# Why the unrestricted weighted least squares should be routinely reported in medical meta-analyses

T.D. Stanley[1,2], John P.A. Ioannidis[2,3,4,5], Maximilian Maier[6], Hristos Doucouliagos[1], Willem M. Otte[7], and František Bartoš[8]

**Keywords:** meta-analysis, random effects, unrestricted weighted least squares, Cochrane Systematic Reviews

[1] Department of Economics, Deakin University, Melbourne, Australia
[2] Meta-Research Innovation Center at Stanford (METRICS), Stanford University, Stanford, California, USA
[3] Stanford Prevention Research Center, Department of Medicine, Stanford University School of Medicine, Stanford, CA, USA
[4] Department of Biomedical Data Science, Stanford University School of Medicine, Stanford, CA, USA
[5] Department of Epidemiology and Biostatistics, Stanford University School of Medicine, Stanford, CA, USA
[6] Behavioural Sciences Group, Warwick Business School, University of Warwick, Coventry, United Kingdom.
[7] Department of Pediatric Neurology, UMC Utrecht Brain Center, University Medical Center Utrecht, and Utrecht University, Utrecht, Netherlands
[8] Department of Psychological Methods, University of Amsterdam, Amsterdam, Netherlands

Declaration of Interest: None
Funding: None
Data sharing statement: Simulation data and code have been shared as part of this submission. No other data collected or analyzed.

Correspondence: T.D. Stanley; 221 Burwood Highway, Burwood, 3125, Victoria, Australia.
Email: tom.stanley1@deakin.edu.au

ABSTRACT

The unrestricted weighted least squares (UWLS) meta-analysis estimator of mean effect is an alternative to the conventional random-effects model (RE). It is a weighted least squares regression estimator that can be represented as a multiplicative random-effects model. UWLS has been shown to fit medical research better than RE as measured by AIC/BIC widely across Cochrane Database of Systematic Reviews (CDSR). The independence of UWLS's mean and heterogeneity estimators provide small-sample advantages that RE does not possess. Large small-sample biases and uncertainty in RE's heterogeneity variance estimates explain most of RE's relatively poor fit along with RE's boundary problem, where RE's heterogeneity variance is estimated to be zero. We prove that UWLS almost always has superior fit at RE's boundary with uncommon exceptions. UWLS has also been found to have generally superior statistical properties: bias, MSE, and coverage relative to RE across 1,665 simulation designs compiled from four published studies authored by different teams of researchers. A recent study in this journal replicated UWLS's superior goodness of fit widely across both the CDSR and a new set of simulations. Due to reporting and interpretation errors, this recent study calls for the continued use of RE as the default meta-analysis estimator with limited applications of UWLS. We address this recent study's concerns and show why UWLS should be routinely reported in medical meta-analyses.

## 1. Introduction

For decades, there have been two popular established approaches to estimate the mean effect in meta-analysis: fixed and random effects. "A 'fixed effect' model assumes that a single parameter value is common to all studies, and a 'random-effects' model that parameters underlying studies follow some distribution" (p. 137).[1] Fixed effect (FE) does not generalize to populations that might differ from the one sampled; whereas, random effects (RE) may generalize to future studies or other populations if drawn from the same distribution of effect sizes.[1,2] RE explicitly accommodates heterogeneity by estimating an additive heterogeneity variance, which then is used to estimate the mean effect and adjust RE's standard error and confidence intervals for the observed heterogeneity. Because the exact population, interventions, outcome measures, protocols, and methods typically vary from study to study in the social and medical sciences, some heterogeneity is generally considered plausible, and RE is routinely used in meta-analysis despite many caveats.[1-3]

More recently, a third estimator, the unrestricted weighted least squares (UWLS), also accommodates and adjusts for heterogeneity by exploiting the mathematical invariance of weighted least squares' (WLS's) variance-covariance to any positive constant, $\gamma$.[4-5] The UWLS estimator is neither fixed nor traditionally random; it borrows strength from both conventional estimators.[6] RE assumes that this heterogeneity has a constant variance, $\tau^2$, independent of sampling errors and sample sizes. By permitting effect size variances to be proportional to their sampling error variance, UWLS allows the heterogeneity variance to vary across studies and to be correlated with a study's sample size.[4-6] UWLS can be interpreted as a random-effects model that allows heterogeneity variance to vary (inversely) with the study's sample size. UWLS' multiplicative constant, like RE's $\tau^2$, models and accounts for observed excess heterogeneity in its estimated SEs and CIs. UWLS is not a common fixed effect model, but rather a specific type of random-effects model.[1] As a least squares estimator, UWLS need not make specific distributional assumptions, and its estimate of mean effect is independent of its estimate of overdispersion. These attributes often make UWLS more robust and reliable than RE, as RE's estimates of $\tau^2$ are widely known to be biased and highly uncertain in small samples, and this instability is transmitted to RE's mean estimate. Over a thousand simulation designs, all of which employed a conventional random-effects data generating process (DGP), have shown that UWLS's statistical properties are generally as good as and often better than RE's.[4-7] What matters more is which approach better reflects and fits actual medical research findings. In 67,308 Cochrane Database of Systematic Reviews (CDSR) meta-analyses, UWLS typically fits medical research findings better than RE, often substantially so, according to the conventional Akaike and Bayesian information criteria (AIC and BIC).[8-11] UWLS typically outperforms RE in medical meta-analyses, often substantially, in both low and high levels of heterogeneity.[8]

In the next sections, we provide methodological context and clarify how UWLS can be interpreted both as a weighted least squares regression and as a specific type of random-effects model that allows heterogeneity variance to scale inversely with study sample size. We demonstrate, thoroughly, where and why UWLS often outperforms RE as measured by AIC/BIC in the CDSR. This allows us to re-evaluate the recent study by Hong and Reed[12] (hereafter HR), who raised concerns about the findings of Stanley et al.[8] and argued for a more limited role for UWLS. Our analysis reveals several methodological characterizations and reporting issues that affect HR's conclusions. Among others, HR misinterprets a genuine modelling advantage for UWLS as a shortcoming of information criteria, and they misreport and underestimate UWLS' coverage. In the process, we justify why UWLS should be routinely reported in medical meta-analyses.

## 2. Alternative estimators of mean effect

First, we discuss alternative estimators of the mean effect size. In the next section, we discuss the associated models, as well as their assumptions, limitations, and interpretations.

Conventional meta-analysis estimators of mean effect are inverse variance, ${}^{1}/{}_{v_i}$, weighted averages of the effect sizes, $y_i$.

$$\hat{\mu} = \frac{\Sigma \, {}^{y_i}/{}_{v_i}}{\Sigma \, {}^{1}/{}_{v_i}}, \qquad \text{i}= 1, 2, \ldots, \text{k} \tag{1}$$

Estimators differ only in how individual variances, $v_i$, are defined. The estimators' variances will be equal or proportional to the inverse sum of these weights, $1/\Sigma \, {}^{1}/{}_{v_i}$.[2,4-6,8]

FE substitutes each study's reported error variance, $SE_i^2$, for $v_i$, and fixed effect's variance is the inverse sum of these weights, $1/\Sigma \, {}^{1}/{}_{SE_i^2}$. Fixed effect estimators make no allowance for genuine heterogeneity of effects often caused by the many differences that are routinely seen across studies, including, among others, in: the populations sampled, the protocols employed, the interventions executed, or the way outcomes are measured. If FE is inappropriately applied to make potentially heterogeneous generalizations, it will underestimate its SE, the width of its confidence interval, and its type I errors.

In contrast, the random effects model explicitly models *random* heterogeneity, which adds an estimated between-study variance, $\hat{\tau}^2$, to the sampling error variance, making $v_i = SE_i^2 + \hat{\tau}^2$. $\tau^2$ is an overdispersion parameter, typically estimated by the DerSimonian-Laird (DL) method of moments or by some iterative maximum likelihood (ML) procedure.[2,13] However, many methods exist for calculating $\tau^2$ and these may sometimes give disparate results.[14] Moreover, with small datasets, as is typical in medical research, uncertainty of the $\tau^2$ estimates can be very large.[15] Regardless, the heterogeneity variance, $\hat{\tau}^2$, must first (or simultaneously) be estimated to calculate the mean effect, $\hat{\mu}$. This requirement can have notable consequences, especially when the number of studies is small (*e.g.,* five or less, as are 50% of the CDSR). If no overdispersion is found among the reported trial outcomes, conventional random effects truncate $\hat{\tau}^2$ at 0 to avoid negative variances. Otherwise, random-effects' variance, $1/\Sigma\, ^1/_{(SE_i^2 + \hat{\tau}^2)}$, SE, and CIs are larger than FE's as they explicitly account for the heterogeneity variance.

For decades, many methodologists have pointed out problems with random effects. Random-effects estimates often exaggerate effect size relative to other estimators of mean effect when there is publication selection bias (PSB), and they are highly sensitive to the accuracy of the between-study variance estimate, $\hat{\tau}^2$, which is known to be biased and very uncertain in small samples.[1,4-6,13-19] For these and other reasons, we offered an alternative estimator, the unrestricted weighted least squares (UWLS).[4]

UWLS allows individual study variances, $v_i$, to be proportional to their known sampling error variances, $v_i = \hat{\gamma} SE_i^2$. $\hat{\gamma} > 0$ is a multiplicative overdispersion constant, which accommodates the heterogeneity found across studies in a meta-analysis. $\hat{\gamma} > 1$ signifies overdispersion or excess between-study heterogeneity, and $\hat{\gamma} < 1$ measures the amount of underdispersion, which RE does not recognize. With $v_i = \hat{\gamma} SE_i^2$, Eq. (1) contains $1/\hat{\gamma} > 0$ in both the numerator and the denominator, which 'cancels out,' making UWLS's estimate of the mean equal to FE's.

$$\hat{\mu}_{UWLS} = \frac{\sum y_i/\hat{\gamma}SE_i^2}{\sum 1/\hat{\gamma}SE_i^2}. = \frac{\sum y_i/SE_i^2}{\sum 1/SE_i^2} = \hat{\mu}_{FE}. \tag{2}$$

The difference is that UWLS does *not* assume a common effect and will automatically accommodate observed between-study heterogeneity, when present, in its estimated variance, $\hat{\gamma}/\Sigma\, ^1/_{SE_i^2}$. How does UWLS do so? The invariance of weighted least squares estimators to any positive multiplicative variance-covariance constant is a well-known property of WLS.[4-6,21,22]

This same UWLS approach has been shown to apply to meta-regressions with any number of moderator variables.[5] When there are no moderators, $\hat{\mu}_{UWLS}$ becomes an inverse-variance weighted average with: weights = $^1/_{SE_i^2}$, variance $= \hat{\gamma}/\Sigma\, ^1/_{SE_i^2}$, and $\hat{\gamma}$ equal to $\hat{\mu}_{UWLS}$'s weighted mean squared errors:

$$\hat{\gamma} = \sum[(y_i - \hat{\mu}_{UWLS})^2/SE_i^2] \Big/ (k-1) \tag{3}$$

In practice, the easiest way to calculate UWLS is from the unweighted ordinary least squares regression of the standardized effect size, $t_i = y_i/SE_i$, on precision, $1/SE_i$, as the only independent variable and with no intercept.[4-6] All statistical software will automatically calculate $\hat{\mu}_{UWLS}$ as the estimated regression coefficient along with its standard error, *t*-test statistic, and CI.[i] This simple regression is also known as the Galbraith radial plot, historically used to represent clinical trial results.[23,24] The mean squared error of this regression is also $\hat{\gamma}$.

$$\hat{\gamma} = \sum\left[t_i - \left(\frac{\hat{\mu}_{UWLS}}{SE_i}\right)\right]^2 \Big/ (k-1) \tag{4}$$

This multiplicative constant, $\hat{\gamma}$, has been called '$H^2$' and considered along with $I^2$ as the "favoured measures for quantifying heterogeneity in a meta-analysis" (p.1553), that is, favored over $\hat{\tau}^2$.[25] $\sqrt{\hat{\gamma}}$ is the ratio of UWLS's SE to FE's SE and "may be interpreted approximately as the ratio of the confidence interval widths for single summary estimates for random effects and fixed effect meta-analyses" (p.1553).[25]

The mathematical relationship between $\hat{\gamma}$ and Cochran's Q also links $\hat{\gamma}$ and $\hat{\tau}^2$. The sum of squared errors of UWLS's regression, $\hat{\gamma}$(k-1), is the well-known Cochran's $Q$ statistic, often used to test for heterogeneity. Because $\gamma = E(Q)/(k-1)$, Higgins and Thompson's[25] Eq.(1) mathematically links these two overdispersion parameters, $\gamma$ and $\tau^2$,

$$\tau^2 = (k-1)(\gamma-1) \Big/ [\Sigma\, 1\!/_{SE_i^2} - \frac{\Sigma 1/_{SE_i^4}}{\Sigma 1/_{SE_i^2}}] \tag{5}$$

Thus, there is a direct connection between these measures of overdispersion. Forcing $\gamma$ = 1 is tantamount to requiring $\tau^2$ = 0, and vice versa. The DerSimonian-Laird method of moments is derived explicitly from this relationship between these parameters.[13,25]

$$\hat{\tau}_{DL}^2 = \max\{0, (k-1)(\hat{\gamma}-1) \Big/ [\Sigma\, 1\!/_{SE_i^2} - \frac{\Sigma 1/_{SE_i^4}}{\Sigma 1/_{SE_i^2}}]\} \tag{6}$$

Thus, RE's estimate of $\tau^2$ can be calculated as a simple function of $\hat{\gamma}$ and UWLS's weights. UWLS also allows the calculation of prediction intervals that explicitly accommodate a random-effects additive heterogeneity.

Furthermore, as a WLS regression, UWLS's mean, variance, and its CIs can be calculated without explicitly using a second parameter or independently estimating it. Passively allowing WLS to be calculated with $SE_i^2$ on the principal diagonal of the variance-covariance matrix will automatically calculate $\hat{\mu}_{UWLS}$, $\hat{\gamma}$, and all of the other statistics with only the given metanalytic data, $\{y_i; SE_i^2\}$. Note that $\hat{\gamma}$ in Eq. (3) is a function of only $\hat{\mu}_{UWLS}$ and the meta-data. For RE, the additive overdispersion metric, $\hat{\tau}^2$, must be known before $\hat{\mu}_{RE}$ can be calculated. Thus, UWLS is a more parsimonious estimator.

Note that $\hat{\mu}_{UWLS}$ is estimated independently of $\hat{\gamma}$, recall Eq. (2). $\hat{\gamma}$ can be any nonzero constant without affecting UWLS's estimate of the mean. $\hat{\mu}_{UWLS}$ and $\hat{\gamma}$ are independent (see the Appendix).[21,22] Thus, uncertainty in $\hat{\gamma}$ will not affect $\hat{\mu}_{UWLS}$. The same cannot be said of RE and this difference can have profound implications about the relative goodness of fit for these estimators of the mean effect (see the Appendix). Because RE's estimate of the mean is dependent on its estimate of the heterogeneity variance, $\hat{\tau}^2$, known to be biased and highly uncertain in small samples, RE can be markedly sensitive to its estimate of overdispersion.[1,2,13-

[19,26,27] In particular, RE's maximum likelihood (ML) estimate of $\tau^2$ is known to be biased downward and unreliable, especially in small samples,[12,26] and RE's fit as measured by AIC/BIC is vulnerable to its estimate of $\tau^2$. In contrast, UWLS's estimate of the mean effect is independent of its estimate of overdispersion, $\hat{\gamma}$, thereby avoiding this basis for RE's uncertainty and unreliability, especially as seen in small samples which are a very common occurrence in biomedical research and many other scientific fields.

RE's relatively poor fit as measured by AIC/BIC is quite striking at its boundary (i.e., $\hat{\tau}^2$=0). When the estimated $\tau^2$ is 0, which happens in 44% of the meta-analyses in the CDSR, RE reduces to the fixed effect estimate but its AIC/BIC still incurs the added parameter penalty. Although UWLS has the exact same penalties as RE in both AIC and BIC, its maximum log likelihood is almost always better because $\hat{\gamma}$ is not constrained to be 1 or greater. Allowing underdispersion and thereby avoiding any boundary condition is a genuine modelling advantage for UWLS. It is not a structural flaw of information criteria nor a small-sample problem. As we prove in the Appendix, UWLS's AIC/BIC will always be better than RE's, regardless of the sample size, whenever $\hat{\tau}^2$=0 and $\hat{\gamma} \neq 1$.

For $\tau^2 > 0$, RE's poor estimation of $\tau^2$ creates a second structural disadvantage in statistical fit. The weighted sum of squared errors (WSSE) is itself a statistical measure of goodness of fit, as it is what weighted least squares minimize. WSSE is also a term in these estimators' log-likelihoods that enters with a negative sign (i.e., a larger WSSE lowers the log-likelihood and worsens the fit). For RE, $WSSE_{RE} = \sum[(y_i - \hat{\mu}_{RE})^2/(SE_i^2 + \hat{\tau}^2)]$. Each term has expected value 1 when the true parameters are substituted for RE's estimators. However, a structural problem occurs in $WSSE_{RE}$ if $\hat{\tau}^2$ is biased downward, which is typical of the ML estimator for $\tau^2 > 0$.[26] $\hat{\tau}^2$'s downward bias makes the denominator too small, on average, and each term of the summation is greater than 1 in expectation. Thus, $E[WSSE_{RE}] > k$ when $\tau^2 > 0$. For UWLS, $WSSE_{UWLS} = \sum[(y_i - \hat{\mu}_{UWLS})^2/\hat{\gamma}SE_i^2]$. The ML estimate of $\gamma$ is $\hat{\gamma} = Q/k = \sum[(y_i - \hat{\mu}_{UWLS})^2/SE_i^2]/k$. Thus, $WSSE_{UWLS} = \sum[(y_i - \hat{\mu}_{UWLS})^2/\hat{\gamma}SE_i^2] = Q/\hat{\gamma} = k$.[ii] This equality holds exactly, regardless of $\hat{\gamma}$'s estimation errors or uncertainty, which is another consequence of the independence of UWLS's estimators—see the Appendix. The same cannot be said of $WSSE_{RE}$. Although random sampling variation means that $WSSE_{RE}$ will not exceed $k$ in every meta-analysis, it will systematically tend to do so whenever $\hat{\tau}^2$ is biased downward. This provides a further reason, beyond the Heterogeneity Boundary Theorem (Appendix), to expect UWLS to often achieve a better statistical fit than RE, especially in small samples, whether statistical fit is measured by log-likelihoods or as some function of the sum of precision-weighted squares.

What else do we know about these alternative estimators of the mean effect? The FE estimator may have some limited applicability if we have a priori knowledge of a common effect and if we wish to make only a limited generalization.[2] Otherwise, the FE estimator will have higher type I errors and poor coverage when there is genuine heterogeneity relative to either RE or UWLS. Thus, we focus on the comparison between RE and UWLS.

First to the obvious, RE weights smaller studies (larger SEs) relatively more heavily than UWLS. If smaller trials tend to be of lower quality or if there is any selection or preference for statistical significance (PSB), then RE will be inferior to UWLS, and PSB has been widely documented to be very common across medicine.[28-33] If, on the other hand, large trials with high precision also have notably different findings that are suspected to be errors or mistaken, then UWLS will be more susceptible to such errors. However, one may see this sensitivity to highly precise yet notably different effect sizes (i.e., influence points) as a potential advantage. With UWLS, suspicious, yet highly influential, effect sizes are more easily identified, forcing the systematic reviewer to carefully investigate the details of these specific trials and the recorded results. In our experience, we have often identified and then corrected coding or reporting errors by looking for precise effect sizes that are notably different from the rest of the results.

Diverse simulation designs, across several different research teams have shown that UWLS's statistical properties are as good or better than RE's.[4-7,18] When there is no PSB, these two estimators have practically equivalent statistical properties: bias, MSE, and coverage rates. When there is PSB, UWLS outperforms RE. It is especially important to compare UWLS to RE using a standard, widely accepted set of simulations. Doing so removes the temptation, whether intentional or unintentional, of choosing designs that favor the authors' preferred methods. For evaluating new PSB methods, Bartoš *et al.*[7] compiled 1,665 simulation designs from Hong and Reed,[34] which included simulations found in four papers authored by four different research teams. Across these 1,665 simulation designs with some assuming no PSB and others allowing different levels of PSB, UWLS (WLS) generally has lower bias, MSE, and better coverage than RE (RMA), see (https://fbartos.github.io/PublicationBiasBenchmark/articles/Results.html).[7]

Giving more weight to precise and high-powered studies is statistically justified as these trials contain more reliable information. Some have gone as far as to recommend that systematic reviewers focus on only the largest clinical trial to the exclusion of all others.[35] These 1,665 publication bias benchmark simulations show that calculating UWLS using only estimates with retrospective power $\geq$ 80% (called the "weighted average of the adequately powered" or WAAP-WLS) has notably superior statistical properties to RE, overall, with better RMSE, bias, coverage, and type I errors.[7,36] Note that WAAP-WLS reverts to UWLS if the number of adequately powered studies is less than 2; thus, it can be calculated for all meta-analyses. WAAP-WLS is especially resilient to PSB as it has the best bias ranking even relative to methods designed to accommodate and correct for PSB. Not only is UWLS's heavy weighting of larger studies relative to RE statistically justified, WAAP-WLS's still heavier weighting has been validated across these 1,665 simulation designs.[7,36]

Thus far, we have only discussed alternative *estimators* of mean effect, not their underpinning models. We turn next to these models.

## 3. Meta-analysis models

"Essentially, all models are wrong, but some are useful" (Box and Draper, 1987, p.424).[37] The utility of models is revealed through the successful applications of their associated estimators. Nonetheless, models provide the mathematical representations of estimators used to derive their theoretical statistical properties.

All three meta-analysis models can be generally represented as:

$$y_i = \mu + e_i; \quad e_i \sim \mathrm{N}(0, v_i) \qquad \text{for } i = 1, 2, \ldots, k; \tag{7}$$

When the variances are known, the respective inverse-variance weighted averages will have the classically desirable statistical properties. When variances are consistently estimated, $SE_i^2$, the respective estimators will have these desirable statistical properties in large samples (i.e., asymptotically).

The fixed effect model assumes that each study's effect size is randomly drawn from the same normally distributed population with a common effect ($\mu$) and only sampling errors to cause study-to-study variation. Sampling error variance, $v_i = \sigma_i^2$, is estimated in the individual studies by $SE_i^2$. Because FE does not allow for heterogeneity of effects, its inferences are limited to this common population.

Random-effects models explicitly accommodate heterogeneity. The conventional RE additive model of heterogeneity sets $v_i = (\sigma_i^2 + \tau^2)$. Or,

$$y_i = \mu + \theta_i + \varepsilon_i, \quad \theta_i \sim N(0,\tau^2),\ \varepsilon_i \sim N(0,\sigma_i^2),\ \text{and } i = 1, 2, \ldots, k; \tag{8}$$

where $\theta_i$ represents an independent and normally distributed heterogeneity.

The multiplicative UWLS model employs an overdispersion parameter, $\gamma$, with $v_i$=$(\gamma\sigma_i^2)$, accommodating excess heterogeneity.[5,6,38-39] This multiplicative overdispersion factor, $\gamma$, has also been interpreted as an intra-class correlation parameter and shown to improve coverage in network meta-analyses of rare events.[39,40] It allows heterogeneity variability to be correlated (inversely) with sample size through $\sigma_i^2$. When smaller studies tend to have greater between-study heterogeneity, $\tau_i^2$ and $\sigma_i^2$ will be correlated, and the multiplicative model will approximate this relationship. Surveys of thousands of meta-analyses in psychology and medical research have found evidence of this correlation, and simulations have shown that UWLS has better statistical properties than RE when such a correlation is present.[6,27]

Without loss of generality, UWLS's variance may be expressed as: $v_i$ =$(\gamma - 1)\sigma_i^2 + \sigma_i^2$, where $(\gamma - 1)\sigma_i^2$ is the variance of the additive heterogeneity, $\theta_i$. Conventional RE assumes fixed heterogeneity, while UWLS allows it to vary inversely with a study's sample size and directly with the sampling error variance. In other words, UWLS's multiplicative model is equivalent to the additive random-effects model, Eq. (8), but with $\theta_i \sim N(0,\ \tau_i^2 = (\gamma - 1)\sigma_i^2)$.

To evaluate goodness of fit across 67,308 Cochrane Database of Systematic Reviews (CDSR) meta-analyses by conventional Akaike and Bayesian information criteria (AIC/BIC), both RE and UWLS were treated, equally, as two parameter models. Yet, 79.4% of 67,308 CDSR MAs found that UWLS fit medical and health research results better than RE, and the odds ratio that a Cochrane systematic review would substantially favor UWLS over RE was 9.33 ($CI_{95\%}$: 8.94; 9.73).[8] UWLS outperformed RE across all subgroups of $\tau^2$and $\gamma$. However, the odds ratio that a Cochrane systematic review would *substantially favor* UWLS over RE was higher than 1 in all subgroups of $\tau^2$and of $\gamma$, except for those with $\gamma$ between 1 and 2. Corollary 2 of the Appendix identifies specific values within (or near) this range where UWLS does not necessarily fit effect sizes better than RE when RE is at its boundary and the WLS, rather than the ML) estimator of $\gamma$ is used. Specifically, the range where RE has an advantage over UWLS at its boundary depends on k, the number of studies in the meta-analysis, and whether the WLS estimate of $\gamma$ is greater than 1 but less than 2.144, 1.539 or 1.230, for k=3, 5, or 10, respectively. This happens in 0.2% of the HR's RE simulations (that is, in 31 of these 13,037 boundary cases). Outside the boundary, UWLS is preferred over RE by 2 to 1 (67%) when RE is the DGP.

These results have recently been independently replicated in 120 nonoverlapping network meta-analyses, where 73% favored UWLS by AIC, and the odds ratio that UWLS is substantially favored ($|\Delta AIC| > 3$) over RE is 2.69.[11] Therefore, statistical theory, simulations, and the findings from tens of thousands of medical meta-analyses highlight the statistical advantages of the UWLS estimator of mean effect, providing a solid foundation upon which to recommend that UWLS *estimates* be routinely reported in medical meta-analyses.

We prefer to model and interpret UWLS in a WLS regression context where specific distributional assumptions are not required, see Stanley and Doucouliagos[5]. Without assuming normality and as long as the variance is known up to a proportional constant, the Gauss-Markov theorem proves that UWLS is the best (minimum variance) linear unbiased estimator,.[4,5,20-22]

For decades, particle physicists have employed a modified version of UWLS to establish the mass and charge of fundamental particles (*e.g.,* neutrinos, electrons, quarks, etc.).[41,42] Yet, they do not explicitly assume the multiplicative model nor invoke the variance proportionality argument. Some fundamental particles' magnitudes are known very precisely; for example, the mass of an electron is known to within a factor of $10^{-9}$. It is also interesting to note that particle physicists often omit measurements with low precision to calculate UWLS on the more precise measurements (p.14),[41] similar to WAAP.

In general, we are dubious that estimated effect sizes calculated from individual subject/patient outcomes will follow a strict multiplicative model, but then no model of random effects, including the conventional RE model is likely to be true.[1] Neither the additive nor the multiplicative variance structure has a privileged claim on representativeness, and there is no logical reason to expect heterogeneity to be additive any more than multiplicative. The RE model's additive variance structure is a modeling convention, adopted largely for computational tractability.[1] The UWLS multiplicative model's proportional scaling with sampling variance is an equally plausible starting point that has practically important implications when heterogeneity is in fact inversely correlated with sample size.[6,11,27]

Beyond the variance structure, effect sizes, as estimates, will not strictly follow a normal distribution in finite samples. At best, normality holds asymptotically. More fundamentally, heterogeneity is often likely to be systematic rather than randomly normal. Studies have different 'true' effects largely because they have systematic differences in population characteristics, interventions, outcome measures, and research context. In practice, most of the systematic variation of effect size will remain unknown because meta-analysis samples sizes are so small (especially in the CDSR) that reliable meta-regression is not a practical option. Observed effect size distributions are typically quite different than the assumed normal and often considerably skewed.[43] Thus, the true DGP is likely to be more complex than either RE's additive or UWLS's multiplicative model.

If there are intra-class correlations across arms of clinical trials, the multiplicative model is further justified.[39] If heterogeneity is higher among small studies than large studies, UWLS will often provide an adequate approximation with better statistical properties than RE.[6,43] Smaller studies may often be of lower quality with higher risk of bias, and less reliability generates higher heterogeneity.[43,44] Early smaller studies may still be perfecting treatment and intervention protocols while later and larger trials will tend to implement more standard and proven treatments.[44] Specification of analyses may be more lax in smaller studies. Even minor data dredging or questionable research practices might more easily distort results, and small studies are easier to disappear and thereby be more selectively reported. Although, in practice, clinical data is unlikely to be generated by a strictly multiplicative model, correlated heterogeneity and/or experimental arms will often approximate a multiplicative model. Like particle physics, we prefer to take a pragmatic and agnostic view, calculating the UWLS estimator without explicitly assuming any model or proportional heterogeneity variance. Even when the random-effects model is correct and there is no PSB, the UWLS and RE estimators have practically equivalent statistical properties.[4-7,36] If there is PSB, UWLS is better than RE, often notably. Decades of experience have shown that weighted least squares are, in general, more robust and resilient than the maximum likelihood methods that RE employs, especially in small samples and when the distributional assumptions are unlikely to be true. After investigating thousands of CDSR meta-analyses, IntHout *et al*.[27] conclude: "The large imprecision with which $\tau^2$ is estimated in a typical small-studies' meta-analysis is another reason for concern, and sensitivity analyses are recommended" (p. 861). UWLS provides an alternative to RE that is not susceptible to the unreliability of its estimated heterogeneity.

## 4. Methodological Characterizations and Reporting Issues of Hong and Reed[12]

Recently, HR reinvestigate whether "Unrestricted weighted least squares represent medical research better than random effects in 67,308 Cochrane meta-analyses."[8] They *successfully* replicate all of Stanley *et al*.'s[8] claims. However, HR use three lines of arguments to largely dismiss UWLS's notably better fit of the CDSR. The first is that information criteria (AIC and BIC) are unreliable, especially in small samples, and should not be used to detect the underlying data generating process (DGP). The second uses erroneous calculations of UWLS's confidence intervals (CI) which result in artificially better CI coverage of RE than UWLS. Third, HR mischaracterizes UWLS as a FE (or common effect) model, which they then use to limit the range of UWLS applications.

### 4.1 Can AIC/BIC be used to compare goodness of fit across 67,308 Cochrane meta-analyses?

Stanley *et al*.[8] offer a meta-science survey of how well alternative meta-analysis estimators statistically fit the effect sizes reported in the 67,308 meta-analyses reported in the CDSR. This meta-science study did not suggest that these information criteria should be used to choose which method to use for any specific meta-analysis (MA), especially as the typical sample size is only 5 studies per meta-analysis. Much like tests for PSB, any goodness-of-fit metric will have low power to discriminate among estimators in small samples. Although individually unreliable in small samples, when aggregated over tens of thousands of areas of medical research and hundreds of thousands of studies/clinical trials, AIC/BIC can provide a practical descriptive indicator of the overall goodness of fit of different MA methods.

AIC and BIC were calculated for RE, FE, and UWLS, for each meta-analysis. Then, the estimator that had the better statistical fit (AIC/BIC) was recorded for each pair-wise comparison, one meta-analysis at a time, along with whether this fit was 'substantially' better ($|\Delta AIC|$ and $|\Delta BIC| > 2$). Next, these simple dichotomous outcomes were aggregated across 67,308 Cochrane MAs to calculate proportions and odds ratios.[8]

HR's simulations assume a three-parameter model, defined by Eq. (8) with $v_i = \gamma(\sigma_i^2 + \tau^2)$ and two overdispersion parameters, $\gamma$ & $\tau^2$. In HR's publicly posted code, combinations of specific values of these parameters along with three small sample sizes = {3, 5, 10} and different values of mean effect generate 180 design combinations, each simulated 1,000 times. HR's central argument is that AIC/BIC often does not correctly identify the random-effects' DGP in small samples relative to three other possible models. However, AIC/BIC's performance is much better for FE and UWLS (called by HR 'UWLS-FE'). HR claims that this shows that the AIC/BIC metrics are unreliable and uses this misinterpretation to dismiss Stanley *et al*.[8] relative assessments of goodness of fit across the CDSR.

HR's Table 5 reports average AIC/BICs across thousands of their simulated MAs. In all cases, both average goodness of fit measures favor UWLS over RE even when the exact RE model with known population sampling error variances is imposed on the simulations' DGP. The fact that UWLS generally fits HR's simulations better than RE in small samples is quite revealing. This is exactly as one would expect knowing the widely accepted problems that the conventional RE *estimator* has in small samples where RE's maximum likelihood (ML) estimate of $\tau^2$ is known to be biased downward and unreliable.[1,2,13-19,26,27,45] As we discussed in Section 2, RE's estimator of mean effect is dependent on its estimate of overdispersion, while UWLS is not. Unreliability and uncertainty in estimating $\tau^2$ will be transmitted to RE's estimate of the mean effect.

RE estimator is especially disadvantaged in the small samples used in HR's simulations, due to its boundary problem, (i.e., truncating $\tau^2$ at 0). The combination of large uncertainty and downward bias of the ML estimate of $\tau^2$ causes it to be often set to zero.[45] In 54% of HR's simulations of the random-effects model, the estimated $\tau^2$ is truncated at 0 even though the true $\tau^2$ is set either at 0.01 or 0.10. Similar truncation is seen in the CDSR where 44% of RE's estimated $\tau^2s$ are zero. When estimated $\tau^2$ is set at 0, UWLS will almost always have better goodness of fit. As we show in the Appendix when $\hat{\tau}^2$=0 and $\hat{\gamma} \neq 1$, UWLS's AIC/BIC will always be better than RE's, regardless of the sample size. For those 54% of RE simulation results with $\hat{\tau}^2$=0, UWLS's AIC and BIC are, on average, 2.64 smaller (substantially better) than RE's.

However, the RE model has problems with high heterogeneity as well. We know that the bias in estimating $\tau^2 > 0$ gets larger with larger values of $\tau^2$,[26] which is confirmed by HR's simulations. For example, the two largest values of $\tau^2$={0.5; 1.0} in HR's simulation code (but not reported in HR's paper) produce downward biases in $\hat{\tau}^2$ of {0.199; 0.368, respectively}. Consistent with our discussion about how RE's fit ($WSSE_{RE}$) is expected to be worse than UWLS's ($WSSE_{UWLS}$) when $\hat{\tau}^2$ is biased downward (recall Section 2), the average difference in AIC/BIC is 0.446 in favor of UWLS for these high levels of heterogeneity. RE is a maximum likelihood estimator that has desirable properties only in large samples. Thus, it is not surprising that RE often has inferior goodness of fit in the small-sample sizes simulated by HR.

HR's simulations corroborate the findings of Stanley *et al*.'s meta-science survey.[8] The UWLS *estimator* fits both HR's simulated effect sizes and those found in actual medical research (CDSR) better than RE's estimator. Hong and Reed[12] show that UWLS *estimator* fits simulated medical research better than the RE *estimator* in the typical MA sample sizes seen in the CDSR *even when there is no publication selection bias and the simulated effect sizes are forced to conform entirely to RE's model.*

Furthermore, whether or not AIC/BIC aligns well with a known DGP model is irrelevant to the question of which estimator best fits 67,308 Cochrane MAs. Accurately identifying the unknown and unknowable DGP model is not relevant to this comparison. Perhaps, UWLS is a better estimator of the additive RE model than the conventional RE estimator in small samples? This explanation of HR's simulation findings is consistent with 1,665 simulation designs previously selected by Hong and Reed.[34] Empirical statistical fit is a property of estimators and can be evaluated independently of which processes may or may not have generated the data. In summary, RE suffers from a boundary problem especially in small samples due to the great uncertainty and downward biasedness of RE's ML estimate of $\tau^2$, and RE's estimate of mean effect is entangled with its estimate $\tau^2$. This bias will also make RE's statistical fit predictably worse than UWLS's. In contrast, UWLS has none of these problems; its estimate of mean effect is entirely independent of its estimate of heterogeneity.

Stanley *et al.*'s.[8] survey is a comprehensive empirical evaluation of how well alternative estimators perform as measured by goodness of fit criteria, AIC/BIC. Both AIC and BIC are monotonically increasing functions of the weighted sum of squared errors. Knowing which method better fits medical research as defined by the CDSR is one important criterion upon which to evaluate estimators, regardless of which DGP model may or may not be operating.

Other medical research statisticians agree, and several have found that the UWLS estimator generally fits medical research better than RE.[8,11,46]

To provide guidance to meta-analysts, it is sufficient to show how well different estimators work in practice. Correctly identifying the DGP is both impossible, in practice, and irrelevant.

**4.2 Which estimator has better coverage?**

HR's second line of argument is that RE has better coverage than UWLS; however, issues of coverage are irrelevant to Stanley *et al*.'s[8] findings about the relative goodness of fit across the CDSR or its conclusions. Nonetheless, poor coverage rates can be an important issue on their own. Most critically, if poor coverage leads to inflated type I errors, this might lead to the mistaken adoption of ineffective medical treatments with potentially notable risks of serious side effects and high costs. But does UWLS have elevated type I errors in practice? Across 67,308 Cochrane meta-analyses, 42.4% of RE's estimates are statistically significant while only 33.9% of UWLS's are statistically significant.[8] Thus, it is unlikely that UWLS has inflated type I errors relative to RE in actual medical research applications, as UWLS typically provides the more conservative evaluation of medical research treatments

HR further evaluated the coverage properties of the compared estimators across 180 simulation conditions coded and reported that UWLS coverage rates were smaller than RE in a subset of these conditions. However, HR's reported UWLS's coverage rates are miscalculated. Table 7A reports UWLS's coverage rates use the standard normal critical values ($z$ = 1.96), contrary to the description in HR's manuscript ("Table 7A uses $z$-critical values when calculating confidence intervals for the FE and RE estimators ($t$-critical values are used for the two UWLS estimators)", p.10; n.8)[12]. Using the correct critical value makes a notable difference especially since the critical t-values for df = k-1= {2,4,9} are considerably larger than the standard normal's 1.96 for 95% CIs. This issue can be verified from the publicly posted code, https://osf.io/3wfqe/ , "Part 01. Simulation" program. There they define UWLS's upper and lower CI limits (i.e., "UWLSFEci.ub" and "UWLSFEci.lb") by explicitly using the standard normal's critical values:

```
" summary(UWLSFE)$coefficients[1] +1.96*summary(UWLSFE)$coefficients[2]
summary(UWLSFE)$coefficients[1] -1.96*summary(UWLSFE)$coefficients[2] "
```

For the sake of transparency, we replaced "1.96" in the above lines of code with the correct critical *t*-values for df = {2,4,9} separately when running simulations for k ={3,5,10}. This produces 180 simulation design conditions with 180,000 results because the "Part 01. Simulation" program contains two additional values of $\tau^2$={0.5; 1.0} that are not reported by Hong and Reed[12]. Across all of HR's simulation conditions, RE's average coverage rate is 87.7% and UWLS's is slightly better, 88.9%. We find similar results, RE (87%) and UWLS (88.8%), when CIs are computed across all values of $\tau^2$ in HR's simulation program defined by HR either as the DGP of the FE or the RE model.[iii] When we confine the simulation results to the two values of $\tau^2$ that HR selected to report in Table 7A for what HR defines as RE's DGP, RE's average coverage is slightly better than UWLS's (92.0% vs. 89.1%) after the miscalculation is corrected. Thus, HR's claim that RE has better coverage rates depends on both selective reporting and misreporting.

A more comprehensive statistical evaluation of RE and UWLS can be obtained from a diverse set of simulations originally compiled by Hong and Reed[34] and further extended by Bartoš *et al*.[7] This evaluation shows that UWLS's (WLS) average overall bias, MSE, and coverage rates across Hong and Reed's[34] 1,665 simulations are generally better than RE's (RMA), see (https://fbartos.github.io/PublicationBiasBenchmark/articles/Results.html).[7] RE can have somewhat better coverage rates for some conditions if we rule out PSB.[7,30] But can medical researchers ever rule out PSB, especially if only a few small studies are published? PSB is "one of the strongest findings across the sciences" (p.370)[47], and dozens of surveys have found evidence of publication selection bias widely throughout the sciences, including medical research.[28-33,47-52] Tests for PSB are known to have poor diagnostic performance, especially low power; thus, it would be foolhardy to use them to choose a meta-analysis estimator, especially in small samples. Past studies and simulations have consistently found that when some studies select for statistical significance, the bias, MSE, and the coverage of UWLS's estimator is better than RE's. [4-7,18,36,54]

### 4.3 Mischaracterization of UWLS

HR mischaracterizes UWLS as a common effect (FE) model and calls UWLS 'UWLS-FE' to emphasize this misunderstanding throughout their paper. "Like FE, the UWLS-FE model maintains the assumption of a common true effect" (p. 2).[12] As we have clarified above, UWLS is a type of RE model that has a multiplicative variance structure. This multiplicative structure is mathematically equivalent to an additive random-effects model where the heterogeneity variance is proportional to the inverse of the sample size. Or, UWLS can be seen as a WLS estimator that accounts for between-study heterogeneity by estimating a multiplicative constant that adjusts its SE and CIs accordingly. A common effect model does not have between-study heterogeneity. As we have shown in Eq. (5), RE's $\tau^2$ is a direct function of UWLS's overdispersion parameter, $\gamma$-1, and positive constants. UWLS methods studies have always emphasized how UWLS accounts for between-study heterogeneity but in a different way than the conventional RE additive model.[4-6,8,36] There would be no reason for the common effect model to account for heterogeneity. This misunderstanding allows HR to dismiss UWLS in their concluding sentence as a variety of fixed effect where the choice between UWLS and FE is a matter of judgement about importance of heterogeneity. The consensus view is that the choice of the common effect model and FE is *not* a "judgement about importance of heterogeneity." FE and its common effect interpretation should be used primarily if the reviewer has reason to believe that there is only a single underlying effect or does not wish to generalize beyond the current population.[1,2] UWLS methods papers have *never* assumed that there is a single effect nor that inference must be limited to the current population. [4-6,36]

## 5. Conclusions

The unrestricted weighted least squares (UWLS) is often the better meta-analysis estimator of mean effect. Like conventional random effects (RE), it fully accounts for observed between-study heterogeneity. It can be represented as a multiplicative RE model or, equivalently, as an additive heterogeneity model where the heterogeneity variance is correlated inversely with the study's sample size. It is widely known that RE's estimate of the heterogeneity variance ($\tau^2$) is biased and highly uncertain in small samples, and this bias and uncertainty are transmitted to RE's estimate of the mean effect. Unlike RE, UWLS need not make specific distributional assumptions, and its estimated mean effect is independent of its estimate of overdispersion, thereby avoiding the sources of much of RE's unreliability in small samples. Thus, it is not surprising that RE often has worse goodness of fit in small samples.

Three large surveys of the medical meta-analyses, encompassing the CDSR, have confirmed this advantage of UWLS over RE as measured by goodness-of-fit criteria AIC/BIC.[8,11,46] UWLS's better goodness of fit is further corroborated by Hong and Reed's[12] recent simulations. In HR's simulations, UWLS always has smaller average AIC/BIC than RE, even if RE's model is imposed fully upon the generated effect sizes.

UWLS's superior statistical properties have also been confirmed in diverse simulation designs across several different research teams.[4-7,18,36,54] When we focus on an agnostic set of 1,665 simulation designs compiled from four different research teams and selected by Hong and Reed,[34] UWLS's bias, MSE, coverage rates are generally better than RE's even though the RE model is imposed upon the data generating process—see (https://fbartos.github.io/PublicationBiasBenchmark/articles/Results.html).[7] All of these simulation studies agree: when there is no PSB, UWLS and RE have practically equivalent statistical properties; when some selection for statistical significance is allowed, UWLS is less biased, has smaller MSE and better coverage than RE. In practice, these differences can be huge in applications with important policy implications where RE can be three or more times larger than UWLS.[55] The risks/benefits of the choice between RE and UWLS are notably asymmetric. When RE model's assumptions are correct and there is no publication selection bias (conditions most favorable to RE), UWLS's statistical properties are practically equivalent to RE's and its fit (AIC/BIC) is better in small samples.[12] However, when there is notable publication selection bias, RE properties are often much worse than UWLS's. Thus, the reporting of UWLS incurs little, if any, cost but offers much potential for gain.

Finally, our previous recommendation deserves repeating: "UWLS should be routinely reported along with RE. When both are reported, we suggest that the more conservative estimate of the mean effect should be regarded as the primary research summary" (p.57).[8]

**Appendix: Why UWLS often fits medical research better than RE**

In this Appendix, we explore the differential statistical properties of the random-effect meta-analysis model (RE) and the unrestricted weighted least squares model (UWLS). First, we offer a theorem that explains why information criteria, AIC/BIC, so often substantially favor UWLS over RE in health and medical meta-analyses. This Heterogeneity Boundary Theorem and its corollaries identify the implications of RE boundary problem for these information criteria, largely explaining RE's poor fit of medical research in small-sample meta-analyses. Second, we show that the estimators of UWLS's two parameters are independent, while RE's are not. This has important implications for the small-sample properties of these alternative approaches to meta-analysis and helps to explain why RE's statistical fit of medical research findings is so poor, relative to UWLS.

1. Heterogeneity Boundary Theorem

In the Theorem below, we evaluate information criteria at the MLE values for all estimators. In particular, the MLE of $\gamma$ is $\hat{\gamma} = \sum[(y_i - \hat{\mu})^2/SE_i^2]/k$. in the Corollaries, we examine the information criteria differences when the WLS estimator of $\gamma$ is used. The WLS estimator of $\gamma$ differs from its MLE by a factor of k/(k−1).

First, note the definitions for AIC and BIC. The Akaike information criterion (AIC) = -2LL+2p; where LL is the maximized log likelihood and p is the number of parameters used by the model in question. The Bayesian information criteria (BIC) = -2LL+pln(k) for k studies in the meta-analysis. Both random effects (RE) and the unrestricted weighted least squares (UWLS) are two parameter models, making the penalties for both estimators exactly the same. Thus, the difference in AIC/BIC will be determined entirely by the difference in their log likelihoods: ΔAIC=ΔBIC = -2LLre+2LLuwls. If this difference is positive, then both information criteria will favor UWLS. Or,

-2LLre+2LLuwls > 0 iff -2LLre > -2LLuwls

**Heterogeneity boundary theorem:** For $\hat{\tau}^2$= 0, ΔAIC=ΔBIC $\geq$ 0, with strict inequality holding for $\hat{\gamma} \neq 1$.

ΔAIC=ΔBIC are evaluated at their maximum likelihood estimated values. RE's log likelihood is:

$$\text{LLre} = \left(-\frac{k}{2}\right)\ln(2\pi) - \left(\frac{1}{2}\right)\sum ln(SE_i^2 + \hat{\tau}^2) - \left(\frac{1}{2}\right) \sum[(y_i - \hat{\mu}_{RE})^2/(SE_i^2 + \hat{\tau}^2)] \qquad \text{(A.1)}$$

UWLS's likelihood is:

$$\text{LLuwls} = \left(-\frac{k}{2}\right)\ln(2\pi) - \left(\frac{1}{2}\right)\sum ln(\hat{\gamma}SE_i^2) - \left(\frac{1}{2}\right) \sum[(y_i - \hat{\mu}_{UWLS})^2/\hat{\gamma}SE_i^2] \qquad \text{(A.2)}$$

Next, we multiply these by -2, for consistency with the form of information criteria formulae.

$$\text{A= -2LLre} = k\ln(2\pi) + \sum ln(SE_i^2 + \hat{\tau}^2) + \sum[(y_i - \hat{\mu}_{RE})^2/(SE_i^2 + \hat{\tau}^2)]$$

$$\text{B=-2LLuwls} = k\ln(2\pi) + \sum ln(\hat{\gamma}SE_i^2) + \sum[(y_i - \hat{\mu}_{UWLS})^2/\hat{\gamma}SE_i^2]$$

Therefore, ΔAIC=ΔBIC = A-B. If A-B > 0, ICs favor UWLS. A-B =

$\sum ln(SE_i^2 + \hat{\tau}^2) + \sum[(y_i - \hat{\mu}_{RE})^2/(SE_i^2 + \hat{\tau}^2)] - \sum ln(\hat{\gamma}SE_i^2) - \sum[(y_i - \hat{\mu}_{UWLS})^2/\hat{\gamma}SE_i^2]$,

as the constants, $k\ln(2\pi)$, subtract out. If $\hat{\tau}^2 = 0$, $\hat{\mu}_{RE} = \hat{\mu}_{UWLS} = \hat{\mu}$, and A-B simplifies to:

$$\text{A-B} = \sum ln(SE_i^2) + \sum[(y_i - \hat{\mu})^2/(SE_i^2)] - kln(\hat{\gamma}) - \sum ln(SE_i^2) - \left(\frac{1}{\hat{\gamma}}\right)\sum[(y_i - \hat{\mu})^2/SE_i^2]$$

$$= \sum[(y_i - \hat{\mu})^2/(SE_i^2)] - kln(\hat{\gamma}) - \left(\frac{1}{\hat{\gamma}}\right)\sum[(y_i - \hat{\mu})^2/SE_i^2]$$

Because the ML estimate of $\gamma$ is $\hat{\gamma} = \sum[(y_i - \hat{\mu})^2/SE_i^2]/k$,

$\text{A-B} = k\hat{\gamma} - kln(\hat{\gamma}) - k$ or

$\text{(A-B)/k} = \hat{\gamma} - ln(\hat{\gamma}) - 1$

Because k > 1, A-B > 0 iff $\hat{\gamma} - 1 - ln(\hat{\gamma}) > 0$ iff $ln(\hat{\gamma}) < \hat{\gamma} - 1$ .

When $\hat{\tau}^2 = 0$, the issue of whether AIC/BIC is favorable to UWLS depends on whether $ln(\hat{\gamma}) < \hat{\gamma} - 1$ or equivalently whether $ln(\hat{\gamma}) - \hat{\gamma} + 1 < 0$. This issue is closely related to the well-known "ln inequality" in real analysis that states that $\ln(x) \leq x - 1$ for all $x > 0$.[1,2]

To prove $ln(\hat{\gamma}) < \hat{\gamma} - 1$, we define

$f(\hat{\gamma}) = ln(\hat{\gamma}) - \hat{\gamma} + 1$ implying that

$f'(\hat{\gamma}) = \frac{1}{\hat{\gamma}} - 1$ & $f''(\hat{\gamma}) = -\frac{1}{\hat{\gamma}^2} < 0$ all $\hat{\gamma} > 0$

Thus, $f(\hat{\gamma})$ is strictly concave for all $\hat{\gamma} > 0$, and its maximum value is found by:

$f'(\hat{\gamma}) = \frac{1}{\hat{\gamma}} - 1 = 0 \quad \rightarrow \quad \hat{\gamma} = 1.$

The maximum value of $f(\hat{\gamma})$ at 1 is $f(1) = ln(1) - 1 + 1 = 0$—See Fig 1.Thus, strict concavity implies that $f(\hat{\gamma}) < 0$ for all $\hat{\gamma} > 0$ if $\hat{\gamma} \neq 1$. The proposition, $ln(\hat{\gamma}) < \hat{\gamma} - 1$, is thereby proved for all $\hat{\gamma} > 0$ if $\hat{\gamma} \neq 1$. It follows that ΔAIC=ΔBIC > 0 and that the information criteria will favor UWLS over RE whenever $\hat{\tau}^2 = 0$ as long as $\hat{\gamma} \neq 1$. If $\hat{\gamma} = 1$, the information criterion will find UWLS and RE equally favorable. This equality of AIC/BIC makes perfect sense, as the FE, RE, and UWLS estimators are identical when $\hat{\tau}^2$= 0 & $\hat{\gamma}$ = 1. With $\hat{\gamma}$ = 1, the observed residual variance exactly matches the fixed-effect model's predictions, which is the case where no multiplicative or additive overdispersion is present.

In Corollary 1 and 2, below, we extend the heterogeneity boundary theorem and explore the range of its applicability for the case where the WLS estimator of $\gamma$ is substituted for its MLE in the AIC/BIC formulas. In the above theorem, all ML estimators are used in the information criteria, AIC/BIC, for both RE and UWLS, including MLE $\hat{\gamma} = \frac{\mathrm{Q}}{\mathrm{k}} = \sum[(y_i - \hat{\mu}_{UWLS})^2/SE_i^2]/k$, as AIC/BIC are properties of maximum likelihood estimators. The WLS estimator of $\mu$ is identical to its MLE; however, the WLS version of $\hat{\gamma}$ divides the same weighted sum of squared errors by k-1 to ensure its unbiasedness.[3-6] For WLS, $\hat{\gamma} = \mathrm{Q}/(\mathrm{k}-1) = \sum[(y_i - \hat{\mu}_{UWLS})^2/SE_i^2]/(k-1)$, and it is this WLS estimator of $\gamma$ that is used below.

**Corollary 1:** Asymptotic Equivalence of WLS: For $\hat{\tau}^2$= 0, $\hat{\gamma} \neq 1$ & k→ ∞, ΔAIC=ΔBIC > 0.

We begin with the same line of reasoning and terms as employed in the heterogeneity boundary theorem to prove ΔAIC=ΔBIC > 0, using ΔAIC=ΔBIC = A-B. For the WLS estimator, $\hat{\gamma}$:

A-B= $(k-1)\hat{\gamma} - k ln(\hat{\gamma}) - (k-1) \rightarrow$

(A-B)/k = $\left[\frac{\mathrm{k}-1}{\mathrm{k}}\right]\hat{\gamma} - ln(\hat{\gamma}) - \left[\frac{\mathrm{k}-1}{\mathrm{k}}\right] = \left[\frac{\mathrm{k}-1}{\mathrm{k}}\right](\hat{\gamma}-1) - ln(\hat{\gamma})$ Again, as (A-B)/k > 0 so is ΔAIC=ΔBIC > 0. (A-B)/k > 0 iff

$$ln(\hat{\gamma}) < \left[\frac{\mathrm{k}-1}{\mathrm{k}}\right](\hat{\gamma}-1).$$

For the WLS estimator of $\hat{\gamma}$, we define: g($\hat{\gamma}$) = $ln(\hat{\gamma}) + \left[\frac{\mathrm{k}-1}{\mathrm{k}}\right](1-\hat{\gamma})$. Note that $\left[\frac{\mathrm{k}-1}{\mathrm{k}}\right](1-\hat{\gamma})$ = $-\left[\frac{\mathrm{k}-1}{\mathrm{k}}\right](\hat{\gamma}-1)$.

$$g'(\hat{\gamma}) = \frac{1}{\hat{\gamma}} - \left[\frac{\mathrm{k}-1}{\mathrm{k}}\right] \quad \& \quad g''(\hat{\gamma}) = -\frac{1}{\hat{\gamma}^2} < 0 \text{ all } \hat{\gamma} > 0$$

Thus, g($\hat{\gamma}$) is strictly concave for all $\hat{\gamma} > 0$, just as f($\hat{\gamma}$), above, is. g($\hat{\gamma}$)'s maximum value is found by:

$$g'(\hat{\gamma}) = \frac{1}{\hat{\gamma}} - \left[\frac{\mathrm{k}-1}{\mathrm{k}}\right] = 0 \quad \rightarrow \quad \hat{\gamma} = \left[\frac{\mathrm{k}}{\mathrm{k}-1}\right]$$

Next, we show that g($\hat{\gamma}$)'s maximum value is >0.

$$g\left(\left[\frac{k}{k-1}\right]\right) = ln\left(\left[\frac{k}{k-1}\right]\right) - \left[\frac{1}{k}\right] > 0 \text{ iff}$$

$$ln\left(\left[\frac{k}{k-1}\right]\right) - \left[\frac{1}{k}\right] > 0 \text{ iff } ln\left(\left[\frac{k}{k-1}\right]\right) > \left[\frac{1}{k}\right].$$

(A.3)

To show that g($\hat{\gamma}$)'s maximum value is > 0 for UWLS's estimator, $\hat{\gamma}$, we exponentiate both sides of this last inequality to isolate the two terms in question, $\left[\frac{k}{k-1}\right]$ & $e^{1/k}$. Abramowitz and Stegun[2] 4.2.31, p.70 offers a well-known inequality, $e^x < 1/(1-x) \quad \forall\, x < 1$, that is relevant to these exact terms. Substituting1/k for x into Abramowitz and Stegun[2] 4.2.31gives:

$$e^{1/k} < 1/\left(1-\frac{1}{k}\right) = 1/\left(\frac{k}{k}-\frac{1}{k}\right) = 1/\left(\frac{k-1}{k}\right) = \left[\frac{k}{k-1}\right].$$

Taking the logarithm of each side of the above inequality gives Ineq. (A.3),

$\left[\frac{1}{k}\right] < ln\left(\left[\frac{k}{k-1}\right]\right)$ or $ln\left(\left[\frac{k}{k-1}\right]\right) > \left[\frac{1}{k}\right]$, which is the exact inequality needed to ensure that g($\hat{\gamma}$)'s maximum value is > 0.

To recap, g($\hat{\gamma}$) is strictly concave, but its maximum value is > 0. This implies that there will be values of $\hat{\gamma}$ for which our pivotal inequalities, $ln(\hat{\gamma}) < \left[\frac{k-1}{k}\right](\hat{\gamma}-1)$ and ΔAIC=ΔBIC > 0, will not be true. Another consequence is that equation, g($\hat{\gamma}$)= $ln(\hat{\gamma}) + \left[\frac{k-1}{k}\right](1-\hat{\gamma}) = 0$, will have two roots. Identifying these roots will specify the range restrictions for the heterogeneity boundary theorem when applied to the WLS estimate of $\hat{\gamma}$. The first root is, as before, at $\hat{\gamma}$ = 1, for all k ≥ 2. That is, g(1) = $ln(1) + \left[\frac{k-1}{k}\right] - \left[\frac{k-1}{k}\right]$ = 0. Thus, for finite k, $ln(\hat{\gamma}) < \left[\frac{k-1}{k}\right](\hat{\gamma}-1)$ and ΔAIC=ΔBIC > 0 only if we add qualifiers to the range of $\hat{\gamma}$, the first of which is $\hat{\gamma}$ < 1. Adding the restriction that $\hat{\gamma}$ < 1 will be sufficient to guarantee that $g(\hat{\gamma}) = ln(\hat{\gamma}) + \left[\frac{k-1}{k}\right](1-\hat{\gamma}) < 0$because g($\hat{\gamma}$)'s maximum is at $\hat{\gamma} = \left[\frac{k}{k-1}\right]$ > 1 and strict concavity for $\hat{\gamma} > 0$ means that $\forall\, \hat{\gamma}$: $0 < \hat{\gamma} < 1$, g($\hat{\gamma}$) < g(1)= 0.

Asymptotically, the second root will approach $\hat{\gamma}$ = 1, consistent with the heterogeneity boundary theorem. To see this, note that as k→ ∞, g($\hat{\gamma}$) → f($\hat{\gamma}$) because g($\hat{\gamma}$)= $ln(\hat{\gamma}) + \left[\frac{k-1}{k}\right](1-\hat{\gamma})$ → f($\hat{\gamma}$) = $ln(\hat{\gamma}) - \hat{\gamma} + 1$ as k → ∞. Recall that f($\hat{\gamma}$) has a unique maximum value of 0 at $\hat{\gamma}$ = 1. Thus, Corollary 1 now follows from the Heterogeneity Boundary Theorem, asymptotically (i.e., in large samples).

We can also show that in finite samples AIC/BIC will favor UWLS over RE for $\hat{\tau}^2$= 0 over a substantial portion of possible values of $\hat{\gamma}$, but not all.

**Corollary 2:** Finite sample boundaries of UWLS's AIC/BIC advantage: For $\hat{\tau}^2$= 0 & k > 2 , ΔAIC=ΔBIC > 0 when either $\hat{\gamma} < 1$ or $\hat{\gamma} > 1 + \frac{2}{k} + \frac{8}{3k^2} + \frac{28}{9k^3} + \frac{464}{135k^4}$ .

Corollary 1, above, establishes that: ΔAIC=ΔBIC > 0 iff $ln(\hat{\gamma}) < \left[\frac{\text{k}-1}{\text{k}}\right](\hat{\gamma} - 1)$, $\hat{\gamma} = \left[\frac{\text{k}}{\text{k}-1}\right]$ maximizes g($\hat{\gamma}$) = $ln(\hat{\gamma}) + \left[\frac{\text{k}-1}{\text{k}}\right](1 - \hat{\gamma})$, the maximum value of g($\hat{\gamma}$) > 0, and there can be two roots of the equation, g($\hat{\gamma}$) = $ln(\hat{\gamma}) + \left[\frac{\text{k}-1}{\text{k}}\right](1 - \hat{\gamma}) = 0$. As before, the first root is $\hat{\gamma}$ = 1 with g(1) = 0. Thus, $ln(\hat{\gamma}) < \left[\frac{\text{k}-1}{\text{k}}\right](\hat{\gamma} - 1)$ and ΔAIC=ΔBIC > 0 hold if we add the qualifier $\hat{\gamma}$ < 1. Next, we find the second roots for specific values of k. For example, when k=3, the second root is $\hat{\gamma} = 2.1440$; i.e., g($\hat{\gamma}$ =2.1440) = 0. Because g($\hat{\gamma}$) is strictly concave, g($\hat{\gamma}$) is negative for all values of $\hat{\gamma}$ larger than 2.1440, as 2.440 is larger than the maximum point, $\hat{\gamma} = \frac{\text{k}}{\text{k}-1}$. Thus, when k=3 and $\hat{\gamma} > 2.1440$, $ln(\hat{\gamma}) < \left[\frac{\text{k}-1}{\text{k}}\right](\hat{\gamma} - 1)$ and ΔAIC=ΔBIC > 0. By solving for the second root of g($\hat{\gamma}$) for other values of k, we find that, for example, g($\hat{\gamma}$ =1.539) = 0 for k= 5 and g($\hat{\gamma} = 1.230$) = 0 for k=10. See Fig 2 for a visual summary of these roots for different values of k. Inductively, we see that that second root decreases as k increases. Corollary 1 showed that the second root approaches $\hat{\gamma}$ = 1 as k → ∞. A safe lower limit for the second root that ensures that ΔAIC=ΔBIC > 0 for finite k could provide practical guidance. A fourth-order asymptotic approximation, $1 + \frac{2}{k} + \frac{8}{3k^2} + \frac{28}{9k^3} + \frac{464}{135k^4}$ , provides such a lower limit for $\hat{\gamma}$ > 1. Thus, ΔAIC=ΔBIC > 0 when either $\hat{\gamma} < 1$ or $\hat{\gamma} > 1 + \frac{2}{k} + \frac{8}{3k^2} + \frac{28}{9k^3} + \frac{464}{135k^4}$ .

2. On the independence of $\hat{\mu}_{UWLS}$ and $\hat{\gamma}$

UWLS may be expressed as a multiplicative random-effects model:

$$y_i = \mu + e_i; \quad e_i \sim \text{N}(0, v_i) \;\; \& \;\; v_i = \gamma SE_i^2 \qquad \text{for } i = 1, 2, \ldots, k; \qquad \text{(A.4)}$$

Where $y_i$ is study i's estimated effect size, $SE_i^2$ is its variance, and $\hat{\mu}_{UWLS}= \hat{\mu}$ –see Section 3. Although the independence of $\hat{\mu}_{UWLS}$ and $\hat{\gamma}$ can be proved for UWLS's multiplicative random-effects model, (A.4), it is unnecessary to do so. The independence of UWLS's $\hat{\mu}_{UWLS}$ and $\hat{\gamma}$ is a direct consequence of the fact that UWLS is weighted least squares linear regression. This statistical independence is widely known and often proved as a basic implication of the classical linear regression model. UWLS is the ordinary least squares (OLS) estimator of slope coefficient in the simple linear regression model.[7-9]

$$t_i = y_i/SE_i = \beta(1/SE_i) + \varepsilon_i \qquad ; \varepsilon_i \sim N(0, \sigma^2) \qquad \text{(A.5)}$$

Where $\sigma^2$ is the conventional OLS regression error variance and is represented by $\gamma$ in the UWLS application. See Section 2. Eq. A.5 fulfills all of the requirements of the classical regression model when the individual study variances, $v_i$, are assumed to be a positive multiple of $SE_i^2$ $(i.e., v_i = \gamma SE_i^2)$. The proof is given in nearly all econometric textbooks, for example.[3-6]

The principal hypothesis test in regression analysis is the t-test of whether the independent and dependent variable are statistically related, $H_0: \beta = 0$. $\beta$ represents mean effect size, $\mu$, in the meta-analysis context, and the estimate slope coefficient, $\hat{\beta}$, is $\hat{\mu}_{UWLS}$—see Section 2. The t-tests of $H_0: \beta = 0$ and $H_0: \mu = 0$ are:

$$t_{\hat{\beta}} = \hat{\beta}/\sqrt{(\hat{\sigma}^2/\Sigma\, 1/_{SE_i^2})} \quad \text{\& for UWLS} \quad t_{UWLS} = \hat{\mu}_{UWLS}/\sqrt{(\hat{\gamma}/\Sigma\, 1/_{SE_i^2})} \quad . \qquad \text{(A.6)}$$

$t_{\hat{\beta}}$ with $H_0: \beta = 0$ tests whether the mean effect in a given area of research is zero. The finite-sample independence of the estimator of the regression coefficient, $\hat{\beta}$ ($\hat{\mu}_{UWLS}$), which is in the numerator of Eq. (A.6), and the estimator of the error variance, $\hat{\sigma}^2$ ($\hat{\gamma}$), in the denominator of Eq. (A.6), is fundamental to regression analysis. It is the independence of these estimators that makes the test of $H_0: \beta = 0$ have an exact Student's t-distribution—see for example Davidson & MacKinnon[5], Section 4.4, pp. 138-41. Thus, $\hat{\mu}_{UWLS}$ and $\hat{\gamma}$ are independent when a multiplicative variance structure is assumed.

In contrast, RE's ML estimators, $\hat{\mu}_{RE}$ and $\hat{\tau}^2$, have no known exact finite-sample distribution, and finite-sample independence has not been established. This stands in direct contrast to UWLS, where the classical linear regression structure guarantees exact finite-sample independence and an exact t-distribution for the test statistic.[3-6] The absence of finite independence is seen in RE's mutually dependent score equations for $\mu$ and $\tau^2$

$$\frac{\partial \ell}{\partial \tau^2} = -\frac{1}{2}\sum_{i=1}^{k}\frac{1}{\tau^2+SE_i^2} + \frac{1}{2}\sum_{i=1}^{k}\frac{(y_i-\mu)^2}{(\tau^2+SE_i^2)^2} = 0 \qquad \text{(A.7)}$$

$$\partial\ell/\partial\mu = \Sigma_i \; (y_i - \mu) \; / \; (SE_i^2 + \tau^2) = 0 \qquad \text{(A.8)}$$

See Eq. (A.1) for RE's log likelihood function. Note that each of the above equations, (A.7) & (A.8), have terms of both $\mu$ *and* $\tau^2$, causing a circular dependency that is unavoidable and structural. That is, neither can be solved as a function of the data alone without the other. Replacing $\mu$ with its ML estimate yields:

$$\sum_{i=1}^{k}\frac{\left(y_i-\frac{\Sigma_j \hat{w}_j y_j}{\Sigma_j \hat{w}_j}\right)^2}{(\hat{\tau}^2+SE_i^2)^2} = \sum_{i=1}^{k}\frac{1}{\hat{\tau}^2+SE_i^2} \quad ; \qquad \text{(A.9)}$$

where RE's weights, $\hat{w}_j = \frac{1}{(\hat{\tau}^2+SE_j^2)}$, clearly depends on $\hat{\tau}^2$. Eq. (A.9) cannot be solved for $\tau^2$ as a function of the data alone. The mutual dependency of the score equations necessitates iterative computation. With UWLS, $\hat{\mu}_{UWLS}$ is easily solved from the data alone—recall Eq. (2). Next, $\hat{\gamma} = \Sigma[(y_i - \hat{\mu}_{UWLS})^2/SE_i^2] / (k-1)$ is calculated from the data knowing $\hat{\mu}_{UWLS}$. That is, UWLS's estimators are solved sequentially, starting with the data alone, which is the essence of independence. RE's absence of finite-sample independence is evidenced by the fact that neither of its estimators can be mathematically expressed without the other. $\hat{\mu}_{RE}$ is a weighted average where $\hat{\tau}^2$ is an essential part of the weights, $1/(SE_i^2 + \hat{\tau}^2)$, and there exist no closed-form ML formula for either $\hat{\tau}^2$ or $\hat{\mu}_{RE}$ that involves only the given data.[iv]

Independence gives UWLS additional desirable small-sample properties that RE does not possess. Independence guarantees that variance of $\hat{\mu}_{UWLS}$ will not be affected by the uncertainty of $\hat{\gamma}'$s estimation.[7,10] Eq. 4, Section 2, shows that all $\hat{\gamma}$ terms in $\hat{\mu}_{UWLS}$ divide out; therefore, $\hat{\mu}_{UWLS}$ does not depend on $\hat{\gamma}$ and, likewise, $\hat{\mu}_{UWLS}$'s variance, $\hat{\gamma}/\Sigma \, {}^{1}/_{SE_i^2}$, does not depend upon the estimator, $\hat{\mu}_{UWLS}$, or the uncertainty of its estimation. Granted that it is not intuitive obvious that $\hat{\mu}_{UWLS}$'s variance does not depend on that $\hat{\mu}_{UWLS}$, it follows directly from these estimators' classical regression properties. Estimated regression coefficients are represented by a vector that is geometrically orthogonal to the vector representing the regression errors and hence its estimated error variance. This geometric orthogonality implies these vectors contain non-overlapping information, which is then shown to entail statistical independence.[2-4] Independence means that uncertainty and unreliability in the estimation of $\gamma$ will not be transmitted to UWLS's estimate, while uncertainty and unreliability of estimating $\tau^2$ in small samples affects RE's estimate of the mean because they are not independent.

Also, the independence of UWLS's estimators gives $t_{UWLS} = \hat{\mu}_{UWLS}/(\hat{\gamma}/\Sigma\, {}^{1}/_{SE_i^2})$ an exact Student's t-distribution with df=k-1, recall Eq. (A.6), regardless of how well or poorly $\hat{\gamma}$ is estimated.[3-6] It is this t-statistic that provides the justification UWLS's test for a nonzero mean effect, $H_0: \mu = 0$. That is, the combination sample statistics in Eq. (A.6) will always be distributed as a known t-distribution without any additional terms or complications as long as the conditions for the classical regression model are satisfied, Eq. (A.5) & $v_i = \gamma SE_i^2$.

The RE model has no exact finite-sample distribution for its test of $H_0: \mu = 0$. This is why the normal distribution, z, is typically used. However, normality will hold only asymptotically, in large samples. In contrast, the typical sample sizes that we find in medical research meta-analyses are quite small.

However, RE's $\hat{\mu}_{RE}$ and $\hat{\tau}^2$ are asymptotically independent because the off-diagonal elements of the Fisher information matrix are zero and these parameters are orthogonal in the sense of Cox & Reid (1987).[11] This orthogonality imparts similar properties to RE as UWLS has in finite samples but only asymptotically for RE. This difference means that RE's variance may be inflated up to a factor of $O(1/k)$ due to the uncertainty and unreliability in the estimation of $\hat{\tau}^2$ in small samples.[11,12] With sample size often as small as 3, 4, or 5 in the CDSR, this variance inflation can be notable, as reflected by the information criteria in both the CDSR and Hong and Reed's[13] simulations.

---

[i] In R, the commands are "reg = lm(t ~ 0 + precision)" and "UWLS = as.numeric(reg$coefficients)" after t and precision has been calculated as suggested in text. In STATA, the command is: "regress t precision, noconstant" . UWLS can also be calculated using either program with a regression of the effect size, y, on a constant with inverse variance weights: "reg=lm(y~1, weights=1/se^2)" and "regress y [aweight = Precision_sq];" where Precision_sq= 1/se^2.

[ii] If, instead, we use the WLS estimate of $\gamma$ that divides Q by (k-1), then this weighted sum of squared errors for UWLS is k-1, which is smaller, and makes UWLS's fit even better.

[iii] This assumes only that $\gamma$=1, which is required by HR's 3-parameter model's definitions to generate either the RE or the FE model. For careful readers of Hong and Reed[12], it is important to understand that HR's 3-parameter model's specific values do not correspond to the actual values of $\gamma$ and $\tau^2$as defined by UWLS or the RE model elsewhere. For example, to force HR's model to conform to the RE model, HR set $\gamma$ =1 & $\tau^2$> 0. However, when $\tau^2$> 0, $\gamma$ must be greater than 1 by Eq. (5). As Eq. (5) shows, if $\gamma$ were truly 1, $\tau^2$ must be 0. Similar inconsistencies occur with conventional parameter definitions for other combinations HR's parameters.

[iv] The DerSimonian-Laird method of moments estimator may be seen as an exception. However, it has its own similar problems of downward bias and high uncertainty in small samples. Expert opinion has generally moved towards favoring ML estimators over DerSimonian-Laird's MM.[8,14]

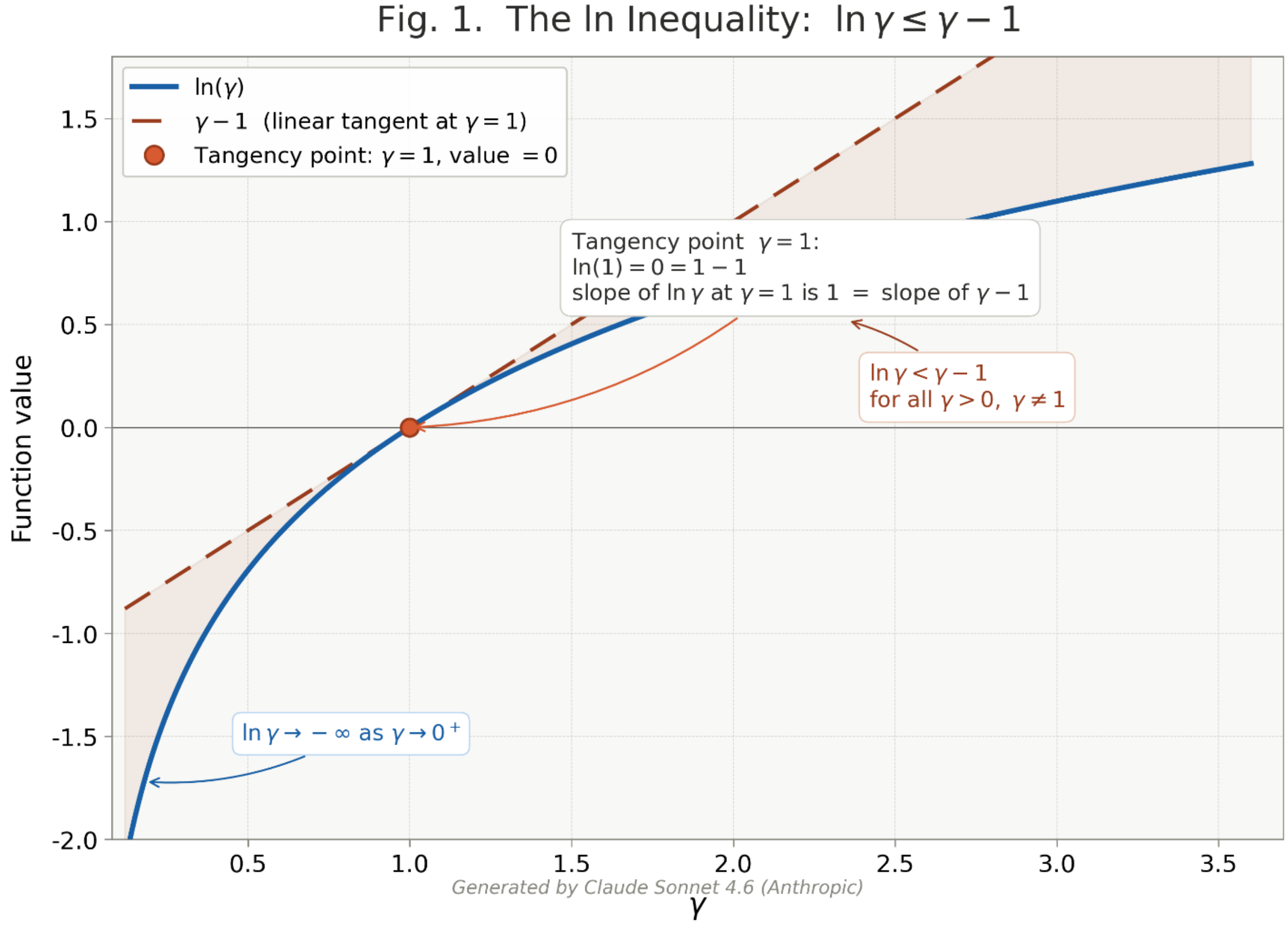
Fig. 1. The ln Inequality: $\ln\gamma \le \gamma - 1$
$\ln(\gamma)$
$\gamma - 1$ (linear tangent at $\gamma = 1$)
Tangency point: $\gamma = 1$, value $= 0$
Tangency point $\gamma = 1$:
$\ln(1) = 0 = 1 - 1$
slope of $\ln\gamma$ at $\gamma = 1$ is $1$ = slope of $\gamma - 1$
$\ln\gamma < \gamma - 1$
for all $\gamma > 0$, $\gamma \neq 1$
$\ln\gamma \to -\infty$ as $\gamma \to 0^+$
Function value
1.5
1.0
0.5
0.0
-0.5
-1.0
-1.5
-2.0
0.5
1.0
1.5
2.0
2.5
3.0
3.5
Generated by Claude Sonnet 4.6 (Anthropic)
$\gamma$

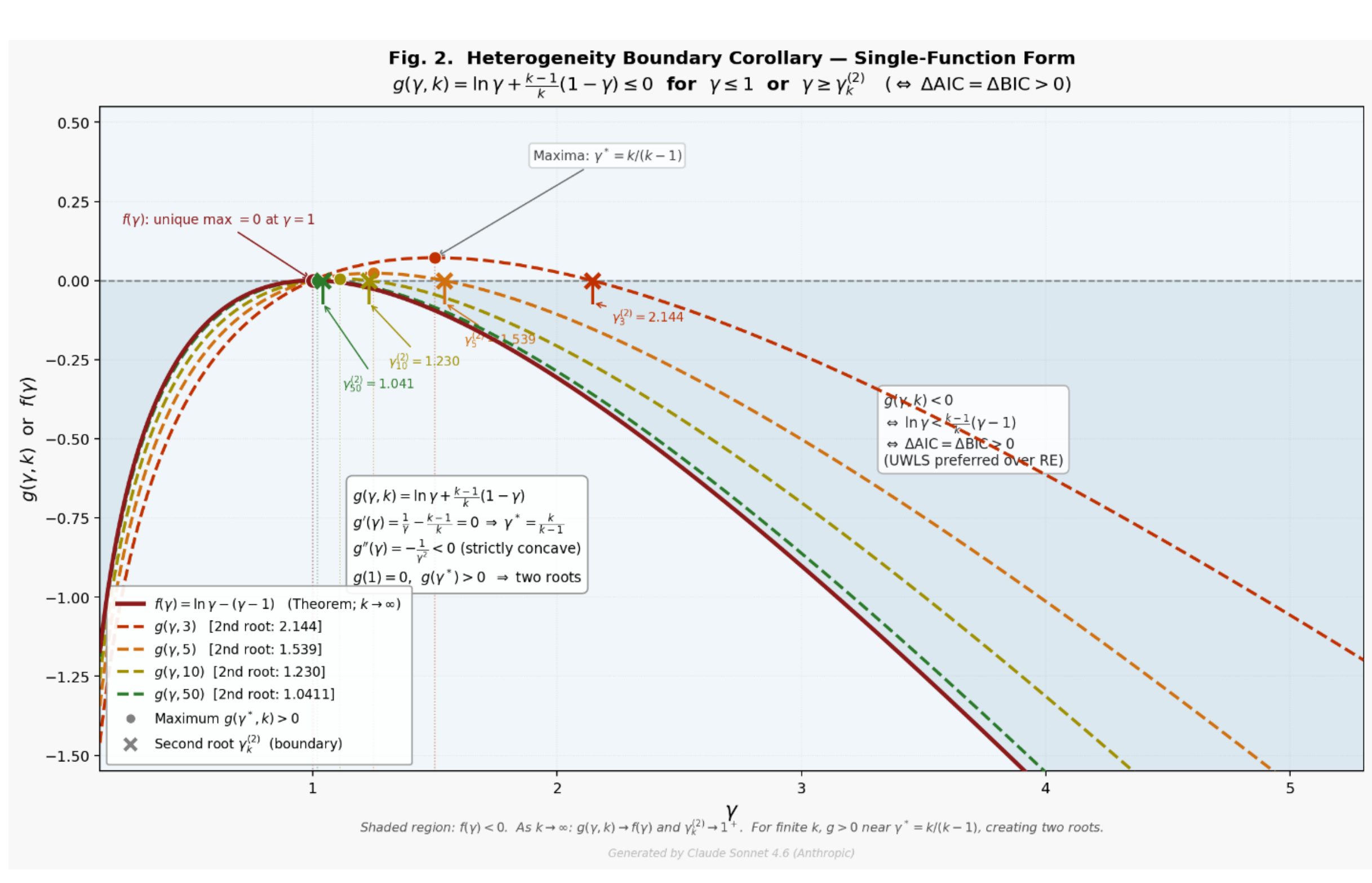
Fig. 2. Heterogeneity Boundary Corollary — Single-Function Form
$g(\gamma, k) = \ln\gamma + \frac{k-1}{k}(1-\gamma) \le 0$ for $\gamma \le 1$ or $\gamma \ge \gamma_k^{(2)}$ ($\Leftrightarrow \Delta AIC = \Delta BIC > 0$)
Maxima: $\gamma^* = k/(k-1)$
$f(\gamma)$: unique max $= 0$ at $\gamma = 1$
$\gamma_3^{(2)} = 2.144$
$\gamma_5^{(2)} = 1.539$
$\gamma_{10}^{(2)} = 1.230$
$\gamma_{50}^{(2)} = 1.041$
$g(\gamma, k) < 0$
$\Leftrightarrow \ln\gamma < \frac{k-1}{k}(\gamma - 1)$
$\Leftrightarrow \Delta AIC = \Delta BIC > 0$
(UWLS preferred over RE)
$g(\gamma, k) = \ln\gamma + \frac{k-1}{k}(1-\gamma)$
$g'(\gamma) = \frac{1}{\gamma} - \frac{k-1}{k} = 0 \Rightarrow \gamma^* = \frac{k}{k-1}$
$g''(\gamma) = -\frac{1}{\gamma^2} < 0$ (strictly concave)
$g(1) = 0$, $g(\gamma^*) > 0 \Rightarrow$ two roots
$f(\gamma) = \ln\gamma - (\gamma - 1)$ (Theorem; $k \to \infty$)
$g(\gamma, 3)$ [2nd root: 2.144]
$g(\gamma, 5)$ [2nd root: 1.539]
$g(\gamma, 10)$ [2nd root: 1.230]
$g(\gamma, 50)$ [2nd root: 1.0411]
Maximum $g(\gamma^*, k) > 0$
Second root $\gamma_k^{(2)}$ (boundary)
$g(\gamma, k)$ or $f(\gamma)$
0.50
0.25
0.00
−0.25
−0.50
−0.75
−1.00
−1.25
−1.50
1
2
3
4
5
$\gamma$
Shaded region: $f(\gamma) < 0$. As $k \to \infty$: $g(\gamma, k) \to f(\gamma)$ and $\gamma_k^{(2)} \to 1^+$. For finite $k$, $g > 0$ near $\gamma^* = k/(k-1)$, creating two roots.
Generated by Claude Sonnet 4.6 (Anthropic)